\documentstyle[aps,twocolumn,epsf]{revtex}

\begin{document}

\draft

\title{
A discrete formulation of teleportation of continuous
variables}
\author{S.J. van Enk}
\address{Norman Bridge Laboratory of Physics, California
Institute of Technology 12-33, Pasadena, CA 91125}

\date{\today}
\maketitle

\begin{abstract}
Teleportation of continuous variables can be described in
two different ways, one in terms of Wigner functions, the
other in terms of discrete basis states. The latter
formulation provides the connection between the theory of
teleportation of continuous degrees of freedom of a light
field and the standard description of teleportation of
discrete variables.
\end{abstract}

Teleportation is a process by which one party, Alice, can transfer any
(unknown) quantum state $|\psi\rangle$ to a distant second party, Bob, by
sending him just the classical information containing the outcome $x_0$ of an
appropriate measurement performed by Alice, provided the two parties share a
nonlocal entangled pair of particles. Alice's measurement is a joint
measurement on two systems, one of which is the particle in the state
$|\psi\rangle$, while the other forms half of the entangled state. Bob can
create the state $|\psi\rangle$ on his part of the entangled state by
applying a unitary operation $U_{x_0}$, the form of which is determined
exclusively by the classical outcome $x_0$. The original protocol of Bennett
{\em et al.} \cite{bennett} concerned quantum states in a finite-dimensional
Hilbert space, so that the measurement outcome $x_0$ is discrete. The
protocol was generalized to continuous variables in \cite{vaidman}. Most
experimental efforts towards accomplishing teleportation using entangled
photons \cite{dik,martini} follow the discrete path.

A recent experiment \cite{akira}, however, succeeded in teleporting
continuous degrees of freedom of a light field, following the theoretical
proposal of Ref.~\cite{sam}. The description of that experiment made use of
the Wigner function, so that its connection to the original teleportation
proposal \cite{bennett} may not be entirely clear. Here we describe the
experiment in the (discrete) photon number state basis and thereby provide
that connection. Moreover, the present formulation is simpler than the one in
Ref.~\cite{stenholm} of teleportation of $N$ variables. The inverse route of
linking continuous to discrete descriptions by reformulating the protocol of
\cite{bennett} in terms of the Wigner function for discrete variables
\cite{wootters} will not be followed here.

In the experiment of Ref.~\cite{akira}, states of a given
single mode \cite{loock} of the electromagnetic field were
teleported. One way of describing the field is in terms of
quadrature amplitudes (see, e.g., Ref.~\cite{walls}), which
are analogous to the position and momentum variables of a
harmonic oscillator (in fact, the electromagnetic field
variables are quantized by first rewriting the Hamiltonian
into the form of an infinite set of harmonic oscillators).
An alternative description is in terms of number states. In
particular, the entangled state that Alice and Bob share is
a two-mode squeezed state, which can be written as
\cite{walls}
\begin{equation}\label{kluns}
|S_r\rangle_{2,3}=\frac{1}{\cosh r}
\sum_{n=0}^{\infty}(\tanh r)^n|n\rangle_2|n\rangle_3,
\end{equation}
where mode $2$ is located in Alice's lab, and mode $3$ in
Bob's. The parameter $r$ is a measure for the amount of
squeezing. The fluctuations in the squeezed variable are
reduced by $\exp(-2r)$, at the cost of increasing the
fluctuations in the complimentary variable by $\exp(2r)$.
For $r\rightarrow\infty$ the two-mode squeezed state is
maximally squeezed and fully entangled. It is interesting
to tabulate the amount of entanglement \cite{ent}, $E=-{\rm
Tr}_2 \rho\log_2\rho$ with $\rho={\rm Tr}_3
|S_r\rangle_{2,3}\langle S_r|$, for a finitely squeezed
state,
\begin{equation}
E=\cosh^2r\log_2(\cosh^2r)-\sinh^2r\log_2(\sinh^2r).
\end{equation}
Figure~1 shows that the amount of entanglement is
approximately linear in the amount of squeezing.
\begin{figure}\label{z} \leavevmode    \epsfxsize=6cm
\epsfbox{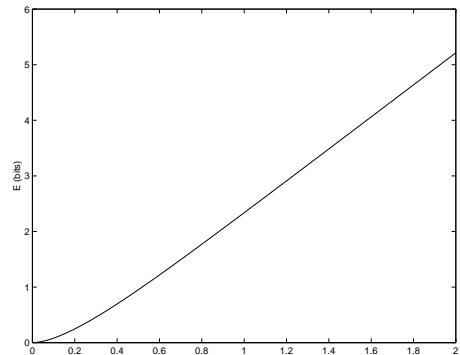}  \caption{Entanglement $E$ in units of bits as a function of
the squeezing parameter $r$. }\end{figure}
 In particular, for $r=0.69$, for
which $\exp(-2r)=0.5$ (the squeezing parameter for the experiment
\cite{akira}), the amount of entanglement is $E=1.46$. The requirements on
the amount of entanglement and corresponding fidelity needed to distinguish
quantum from ``classical'' teleportation, is discussed in \cite{fuchs}.

 Alice is given another field mode
$1$ which is in the state $|\psi\rangle_1$ to be
teleported. This state can be expanded as
\begin{equation}
|\psi\rangle_1=\sum_{n=0}^{\infty}\alpha_n|n\rangle_1.
\end{equation}
As in the original teleportation protocol, Alice has to
perform a joint measurement on modes $1$ and $2$. In
\cite{akira} the joint measurement consisted of two
measurements of the two commuting observables
$\hat{X}=\hat{x}_2-\hat{x}_1$ and
$\hat{P}=(\hat{p}_2+\hat{p}_1)/2$, where $\hat{x}_i$ and
$\hat{p}_i$ are proportional to the quadrature amplitudes
referred to above,
\begin{eqnarray}
\hat{x}_i&=&\frac{1}{2}(a_i+a_i^{\dagger}),\nonumber\\
\hat{p}_i&=&\frac{1}{2i}(a_i-a_i^{\dagger}),
\end{eqnarray}
in terms of the creation and annihilation operators acting
on the modes $i=1,2$. The joint eigenstate of the two
operators $\hat{X}$ and $\hat{P}$ with eigenvalues $X$ and
$P$ can be expanded in the eigenstate basis of $\hat{x}_i$,
\begin{eqnarray}\label{sukkel}
|\phi(X,P)\rangle_{1,2}&=&\int {\rm d}X_1 \int {\rm d}
X_2\, \delta(X_2-X_1-X)\nonumber\\
&&\times\exp(iP(X_1+X_2))|X_1\rangle_1|X_2\rangle_2.
\end{eqnarray}
This state is fully entangled, and is in fact of the same
form as the original EPR state \cite{epr}.

Now in order to discuss the limit of infinite squeezing, in
which the state (\ref{kluns}) is no longer normalizable, we
now truncate the Hilbert space and consider only photon
numbers up to and including $N$, where we may take the
limit $N\rightarrow\infty$ in the end. In particular, the
two-mode squeezed state in the limit of infinite squeezing
$r\rightarrow \infty$ becomes
\begin{equation}
|S_{\infty}\rangle_{2,3}=\frac{1}{\sqrt{N+1}}\sum_{n=0}^N
|n\rangle_2|n\rangle_3.
\end{equation}
We can rewrite the eigenstate (\ref{sukkel}) in that
truncated space as
\begin{equation}\label{froett}
|\phi(X,P)\rangle_{1,2}=\sum_{m=0}^N\sum_{n=0}^N
\gamma_{mn}(X,P)|m\rangle_1|n\rangle_2,
\end{equation}
where we do not yet have to specify the precise form of the
coefficients $\gamma_{mn}(X,P)$ (but see below).  It is
easy to verify that the reduced density matrix of either
mode $1$ or $2$ in the eigenstate (\ref{sukkel}) is
proportional to the identity matrix. This implies that
after Alice's measurement no information  about the
identity of the state $|\psi\rangle$ will be left behind in
either system $1$ or $2$, which is a necessary condition
for faithful teleportation \cite{bennett}. The fact that
\begin{equation}
{\rm Tr}_2 |\phi(X,P)\rangle_{1,2}\langle\phi(X,P)|=
 \frac{1}{N+1}I_1
\end{equation}
with $I_1$ the $(N+1)\times (N+1)$ identity operator on
mode $1$, implies that the coefficients $\gamma_{mn}$
satisfy
\begin{equation}
(N+1)\sum_{l=0}^N\gamma^*_{ml}(X,P)\gamma_{nl}(X,P)=\delta_{mn}.
\end{equation}
That is, the matrix $\sqrt{N+1}\gamma_{mn}$ is unitary. In
order to show explicitly that this is a necessary and
sufficient condition for teleportation to be possible, we
rewrite the joint initial state of modes $1,2,3$, in the
case of infinite squeezing, as
\begin{eqnarray}\label{knar}
|\psi\rangle_1|S_{\infty}\rangle_{2,3}&=&
\frac{1}{N+1}\sum_{X=X_0}^{X_N}\sum_{P=P_0}^{P_N}
\sum_{l=0}^N\sum_{m=0}^N\nonumber\\ &&\gamma_{lm}(X,P)
|l\rangle_1|m\rangle_2\sum_{n=0}^N\beta_n(X,P)|n\rangle_3.
\end{eqnarray}
Here we used that the eigenfunctions of the operators
$\hat{X}$ and $\hat{P}$ form a complete set, so that the
sum ---$P$ and $X$ have become discrete variables now---
over all eigenvalues $X$ and $P$ of the operators
$|\phi(X,P)\rangle_{1,2}\langle \phi(X,P)|$ gives the
identity. The coefficients $\beta_n$ are given by
\begin{equation}
\beta_n(X,P)=\sqrt{N+1}\sum_{m=0}^N
\gamma^*_{mn}(X,P)\alpha_m.
\end{equation}
It follows directly from (\ref{knar}) that after Alice
finds two measurement outcomes $X_0$ and $P_0$, Bob's state
is collapsed onto
\begin{eqnarray}
|\Psi\rangle_3&=&\sum_{n=0}^N\beta_n(X_0,P_0)|n\rangle_3
\nonumber\\
&=&\sqrt{N+1}\sum_{m=0}^N\sum_{n=0}^N\gamma^*_{mn}(X_0,P_0)
\alpha_m|n\rangle_3.
\end{eqnarray}
In order for Bob to be able to recover the original state
$|\psi\rangle$ from $|\Psi\rangle$, we see now that the
matrix $\sqrt{N+1}\gamma_{mn}$ indeed must be unitary: Bob
has to apply the operation
\begin{equation}\label{huppeltrut}
U_{X_0,P_0}:|n\rangle_3\mapsto \sqrt{N+1}\sum_{m=0}^N
\gamma_{mn}|m\rangle_3
\end{equation}
to effect the transformation
\begin{equation}
|\Psi\rangle_3\mapsto|\psi\rangle_3,
\end{equation}
which completes the teleportation process.

 Thus, Bob's unitary operation (\ref{huppeltrut})
and Alice's measurement outcome (\ref{froett}) are both
described by a single unitary matrix $\gamma_{mn}$ (just as
in the example given in \cite{bennett}). In the experiment
\cite{akira} this translates into the fact that Alice's
measurement outcomes are classical currents that Bob
directly converts into field amplitudes and subsequently
mixes with his part of the two-mode squeezed state.

For completeness, let us now calculate the explicit form of
the eigenstates $|\phi(X,P)\rangle_{1,2}$ of the operators
$\hat{X}$ and $\hat{P}$ with eigenvalues $X$ and $P$ in the
number-state basis. First, the (truncated) eigenstate
$|\phi(0,0)\rangle_{1,2}$ with zero eigenvalues is easily
found by simply solving the eigenvalue equations
\begin{eqnarray}\label{Q}
(a_1-a^{\dagger}_2)|\phi(0,0)\rangle_{1,2}&=&0,\nonumber\\
(a_2-a^{\dagger}_1)|\phi(0,0)\rangle_{1,2}&=&0,
\end{eqnarray}
with the result
\begin{equation}
|\phi(0,0)\rangle_{1,2}=\frac{1}{\sqrt{N+1}}\sum_{n=0}^N
|n\rangle_1|n\rangle_2.
\end{equation}
Then, introducing the two commuting operators
$\hat{Y}=(\hat{x}_1+\hat{x}_2)$ and
$\hat{Q}=(\hat{p}_1-\hat{p}_2)/2$, it is easy to verify,
using the commutation relations between $\hat{X}$ and
$\hat{Q}$, and between $\hat{P}$ and $\hat{Y}$, that
\begin{equation}
|\phi(X,P)\rangle_{1,2}=\exp(iP\hat{Y})\exp(iX\hat{Q})
|\phi(0,0)\rangle_{1,2}
\end{equation}
is indeed the desired eigenstate with eigenvalues $X$ and
$P$. Using standard identities for exponentials of creation
and annihilation operators \cite{vogel} and the relations
(\ref{Q}) this state can be rewritten as
\begin{eqnarray}
|\phi(X,P)\rangle_{1,2}&=& \exp(-(P^2+(X/2)^2)/4)
\exp((iP-X/2)a_1)\nonumber\\&& \times
\exp((iP+X/2)a_2)|\phi(0,0)\rangle_{1,2},
\end{eqnarray}
which can be expanded as
\begin{eqnarray}
|\phi(X,P)\rangle_{1,2}&=&
 \exp(-(P^2+(X/2)^2)/4)\nonumber\\
&&\times\sum_{m=0}^N\sum_{n=0}^N\sum_{l={\rm min}(m,n)}^N
\left(\frac{(l!)^2}{m!n!}\right)^{1/2}\nonumber\\
&&\times\frac{(iP-X/2)^{l-m}(iP+X/2)^{l-n}}{(l-m)!(l-n)!}
|m\rangle_1|n\rangle_2.\nonumber\\
\end{eqnarray}
This then yields in the coefficients $\gamma_{mn}$. Because
of the still relatively complicated form of the
coefficients $\gamma_{mn}$, the question how finite
squeezing affects the fidelity of the teleportation process
is better discussed in the Wigner state formalism
\cite{sam}.

In conclusion, the teleportation experiment of
Ref.~\cite{akira} of continuous degrees of freedom of a
light beam can be formulated in the number-state basis,
thus providing a connection with the original formulation
of the teleportation protocol. The measurements of
quadrature amplitudes on Alice's side correspond to
entangled measurements that leave no information behind in
Alice's field modes about the state to be teleported.  This
enables Bob to recreate that state in a field mode in his
laboratory by applying a particular unitary operation,
described by the same unitary matrix $\gamma_{mn}$ that
describes Alice's measurement scheme.

 It is a pleasure to thank C.A. Fuchs, H.J. Kimble
 and especially A. Furusawa
for helpful discussions and for convincing me to finally
publish these notes.
 This work was funded by DARPA through the
QUIC (Quantum Information and Computing) program
administered by the US Army Research Office, the National
Science Foundation, and the Office of Naval Research.

\end{document}